\documentclass[preprint,aps,  nofootinbib]{revtex4}
\usepackage{amssymb}
\usepackage{amsmath}
\usepackage{graphicx}

\newcommand\beq{ \begin{eqnarray} }
\newcommand\eeq{ \end{eqnarray} }

\begin{document}

\title{Shear Viscosity in Weakly Coupled $N$-Component Scalar Field Theories}
\author{Jiunn-Wei Chen$^{1}$, Mei Huang$^{2}$, Chang-Tse Hsieh$^{1}$,
Han-Hsin Lin$^{1}$}
\affiliation{$^{1}$ Department of Physics and Center for Theoretical Sciences, National
Taiwan University, Taipei 10617 }
\affiliation{$^{2}$ Institute of High Energy Physics, Chinese Academy of Sciences,
Beijing 100049}

\begin{abstract}
The rich phenomena of the shear viscosity ($\eta )$ to entropy density ($s)$
ratio, $\eta /s$, in weakly coupled $N$-component scalar field theories are
studied. $\eta /s$ can have a \textquotedblleft double
dip\textquotedblright\ behavior due to resonances and the phase transition.
If an explicit goldstone mass term is added, then $\eta /s$ can either
decrease monotonically in temperature or, as seen in many other systems,
reach a minimum at the phase transition. We also show how to go beyond the
original variational approach to make the Boltzmann equation computation of $%
\eta $ systematic.
\end{abstract}

\maketitle


\section{Introduction}

Scalar field theories are important tools to study spontaneous symmetry
breaking. They can be used to demonstrate the Goldstone theorem by breaking
a global symmetry and the Higgs mechanism by breaking a local symmetry.
Furthermore, in some systems, scalar field theories are the low energy
effective field theories of the underlying theories, such that model
independent results can be obtained. A good example is the universality of
the critical exponents for systems with the same symmetries near second
order phase transitions.

In the study of transport coefficients, scalar field theories also play an
important role. It is proven that in a weakly coupled scalar field theory
with quartic and cubic terms, summing the leading order (in the coupling
constant expansion) diagrams in the Kubo formula for shear viscosity ($\eta $%
) is equivalent to solving the Boltzmann equation with effective temperature
($T$) dependent masses and scattering amplitudes \cite{Jeon}.

Recently, there is a renewed interest in shear viscosity. It was conjectured 
\cite{KOVT1} that no matter how strong the particle interaction is, $\eta /s$
($\eta $ per entropy density) has a universal minimum bound $1/(4\pi )$ in
any system. This bound is motivated by the uncertainty principle and is
found to be saturated for a large class of strongly interacting quantum
field theories whose dual descriptions in string theory involve black holes
in anti-de Sitter space \cite%
{Policastro:2001yc,Policastro:2002se,Herzog:2002fn,Buchel:2003tz}. Much
progress has been made in testing this bound and trying to identify the most
perfect fluid with the smallest $\eta /s$ (see \cite%
{Son:2007vk,Kapusta:2008vb,Schafer:2009dj} for recent reviews). It is found
that $\eta /s$ can be as small as possible \cite{Jakovac:2009xn} (but still
positive) in a carefully engineered meson system \cite%
{Cohen:2007qr,Cherman:2007fj}, although the system is metastable. Also, in
strongly interacting conformal field theories, $1/N_{c}$ corrections, with $%
N_{c}$ the size of the gauge group, can modify the $\eta /s$ bound slightly 
\cite{Kats:2007mq,Brigante:2007nu,Brigante:2008gz,Buchel:2008vz}.

In the real world, the smallest $\eta /s$ known so far belongs to a system
of hot and dense matter thought to be quark gluon plasma ( see \cite%
{Gyulassy:2004zy,Shuryak:2004cy,Stoecker:2004qu,Jacobs:2004qv} for reviews)\
just above the phase transition temperature produced at RHIC \cite%
{RHIC,Huovinen:2001cy,Molnar:2001ux,Teaney:2000cw,Hirano:2002ds,
Teaney:2003pb,Muronga:2004sf,Heinz:2005bw,Romatschke:2007mq}\ with $\eta
/s=0.1\pm 0.1(\mathrm{theory})\pm 0.08(\mathrm{experiment})$ \cite%
{Luzum:2008cw}. A robust upper limit $\eta /s<5\times 1/(4\pi )$ was
extracted by another group \cite{Song:2008hj} and a lattice computation of
gluon plasma yields $\eta /s=0.134(33)$ \cite{etas-gluon-lat}. Progress has
been made in cold unitary fermi gases as well. An analysis of the damping of
collective oscillations gives $\eta /s\gtrsim 0.5$ \cite{Schafer,Turlapov}.
Even smaller values of $\eta /s$ are indicated by recent data on the
expansion of rotating clouds \cite{Clancy,Thomas} but more careful analyses
are needed \cite{Schaefer2}.

In general, stronger interaction implies smaller $\eta /s$ and it is found
that $\eta /s$ goes to a local minimum near the phase transition temperature
($T_{c}$) in a large class of systems \cite%
{Csernai:2006zz,Chen:2006iga,Chen:2007jq}. In particular, $\eta /s$ develops
a cusp(jump) at $T_{c}$ for a second(first) order phase transition and a
smooth local minimum for a cross over. This behavior is seen in QCD with
zero baryon chemical potential \cite%
{Csernai:2006zz,Chen:2006iga,Arnold:2003zc,Xu:2007ns,Chen:2009sm,Prakash:1993bt,Dobado:2003wr,Dobado:2001jf}
and near the nuclear liquid-gas phase transition \cite%
{Chen:2007xe,Itakura:2007mx}. It is also seen in cold unitary fermi gases 
\cite{etas-supfluid}, in H$_{2}$O, N, and He and in all the matters with
data available in the NIST database \cite{webbook,Csernai:2006zz,Chen:2007xe}%
. Thus, it was speculated that this feature is universal. If this is indeed
the case, then $\eta /s$ can be used to probe some parts of the systems
which are hard to explore otherwise. For example, one can try to locate the
critical point of QCD by measuring $\eta /s$ \cite{Lacey:2006bc,Chen:2007xe}.

Theoretically, the behavior of $\eta /s$ going to a local minimum near the
phase transition can be reproduced in controlled calculations of
one-component weakly interacting real scalar field theories \cite%
{Chen:2007jq}. In this paper, we extend the calculation to $N$-component
real scalar field theories (see \cite{Aarts:2004sd,Moore:2007ib} for earlier
work in the symmetric phase and work in linear sigma models with QCD-like
parameters \cite{Dobado:2009ek,Chakraborty:2010fr}). Unlike the one
component case, the resonance (called $\sigma $) in the intermediate states
makes the power counting more complicated. To perform a controlled
calculation, we take the $N\rightarrow \infty $ limit and make the goldstone
bosons (called $\pi $'s) massive to stay away from the $\sigma $ pole. The
resulting $\eta /s$ can decrease monotonically in temperature despite the
phase transition. Also, when the resonance effect is important, $\eta /s$
can have a \textquotedblleft double dip\textquotedblright\ behavior. It is
conceivable that by tuning parameters, one can make the two local minimums
to be close to each other and with the cusp at $T_{c}$\ smoothed out such
that one just sees the single minimum below $T_{c}$ as is shown in \cite%
{Dobado:2009ek}. These are behaviors different from that of the
one-component theory \cite{Chen:2007jq} and many other systems mentioned
above. We also show how to go beyond the original variational approach to
make the Boltzmann equation computation of $\eta $ systematic.

\section{Validity of the Boltzmann Equation}

It was proved first in weakly coupled $\phi ^{4}$ theory \cite{Jeon} and
then in hot QED \cite{Gagnon:2007qt} that the Boltzmann equation gives the
sum of the leading order diagrams in the coupling constant expansion.
Applying this equality to strongly interacting systems is dangerous because
the Boltzmann equation might become invalid. As a semi-classical approach,
the Boltzmann equation describes the evolution of the distribution function $%
f(\mathbf{x},\mathbf{p},t)$, which specifies $\mathbf{x}$ and $\mathbf{p}$
at the same time. Thus, to be consistent with quantum mechanics, the size of
the collision region ($l_{sc}\sim \sigma _{sc}^{1/2}$, with $\sigma _{sc}$
the scattering cross section) needs to be much smaller than the mean free
path ($l_{mfp}\sim 1/n\sigma _{sc}$, with $n$ the number density). In other
words, the quantum mechanical collisions happen in short distance ($l_{sc}$)
while particles travel for long distances ($l_{mfp}$) as free particles
between collisions. It is the wide separation of the two distance scales $%
\left( l_{sc}/l_{mfp}\ll 1\right) $ that allows the semi-classical treatment
of quasi-particle collisions in the Boltzmann equation.

For a theory with a dimensionless coupling constant $g$, if all the other
mass scales are of the same order as the temperature $T$, then $%
l_{sc}/l_{mfp}\sim n\sigma _{sc}^{3/2}\propto g^{\alpha }$, where $\alpha $
is positive for perturbation theories. For example, in a gas of massless
pions, $g\sim T/f_{\pi }$ and $\alpha =6$. Thus the Boltzmann equation is
applicable at low $T$ \cite{Chen:2006iga}. Also, in one-component real scale
field theories \cite{Chen:2007jq}, $\alpha =3$, with $g$ the coupling of the
four-scalar vertex. In strongly interacting systems, in general $%
l_{sc}/l_{mfp}\ll 1$ is not satisfied and the Boltzmann equation is no
longer applicable.

In $N$-component scalar field theories with a spontaneously\ broken $O(N)$
symmetry, the $\sigma $ resonance could make $l_{sc}/l_{mfp}\sim 1$ even
when the coupling is weak. We will first discuss the case with $N=\mathcal{O}%
(1)$ then move to the large $N$ case. In the symmetry breaking phase, the $%
\pi \pi $ scattering amplitude can be enhanced through the s-channel $\sigma 
$ pole in intermediate state. Adding the $\sigma \rightarrow \pi \pi $\
decay width $\Gamma _{\sigma }$ which is $\mathcal{O}(\lambda )$, the cross
section $\sigma _{sc}$ becomes $\mathcal{O}(\lambda ^{0})$ near the $\sigma $
resonance. This gives $l_{sc}/l_{mfp}\sim 1$ and the Boltzmann equation is
no longer applicable.

In the large $N$ case, however, $g$ can be scaled as $1/N$ such that the
one-loop result is the same order in $N$ as the tree-level one. Therefore, $%
l_{sc}\sim g\sim 1/N$, $l_{mfp}\sim 1/(Ng^{2})\sim N$, and $%
l_{sc}/l_{mfp}\sim 1/N^{2}$ despite the $\sigma $ resonance. This is yet
another example that large $N$ systems are classical in nature. However,
even though we will perform such a computation in the large $N$ limit, it is
not clear whether $l_{sc}/l_{mfp}\ll 1$ is sufficient to justify the use of
the Boltzmann equation in the $\eta $ computation.

One way to stay away from the $\sigma $ pole to obtain a reliable $\eta $ is
to add a mass term to $\pi $'s. Because if $2m_{\pi }>m_{\sigma }$, then $%
\sigma $ will never be on-shell in the $\pi \pi $ scattering. We will also
present results in this case.

\section{Modified O($N$) Model in the Large $N$ Limit}

We will study a $N$-components scalar theory with the Lagrangian%
\begin{equation}
\mathcal{L}=\frac{1}{2}(\partial _{\mu }\vec{\phi})^{2}-\frac{1}{2}a\vec{\phi%
}^{2}-\frac{1}{2}m^{2}\vec{\pi}^{2}-\frac{1}{4}\frac{b}{N}\left( \vec{\phi}%
^{2}\right) ^{2},
\end{equation}%
where $\vec{\phi}=(\vec{\pi},\phi _{N})$ and $\vec{\pi}=(\phi _{1},\phi
_{2},\cdots ,\phi _{N-1})$. When $m=0$, this theory has an $O(N)$ symmetry
such that the Lagrangian is invariant under $\phi _{i}\rightarrow R_{ij}\phi
_{j}$ with $R_{ij}R_{ji}=1$. The $m^{2}$ term break the $O(N)$ symmetry to $%
O(N-1)$, thus it is called a modified $O(N)$ model in this paper. As
mentioned above, the inclusion of the $m^{2}$\ term ($m^{2}>0$) is to avoid
the production of on-shell $\sigma $. $a$, $b$ and $m$ are renormalized
quantities and the counterterm Lagrangian is not shown. The renormalization
condition is that,\ at $T=0$, the counterterms do not change the particle
mass and the four-point couplings at threshold. The $1/N$ scaling in the $b$
coupling makes sure that the quasi-particle masses are $\mathcal{O}(N^{0})$.
We will discuss the following cases: (I) $a>0,$ $b>0,$ the system is always
in the symmetric phase. (II) $a<0,$ $b>0,$ in the $m\rightarrow 0$ limit,
the vacuum at $T=0$ breaks the $O(N)$ symmetry spontaneously and there are $%
N-1$ massless goldstone bosons. Those goldstone bosons become massive
because of the $m^{2}$ term. At higher $T$, the symmetry is restored through
a second-order phase transition. (III) Adding a term 
\begin{equation}
\delta L=-\sqrt{N}H\phi _{N}
\end{equation}
to the Lagrangian of (II) to model a crossover.

We will focus on the case of weak coupling in the large $N$ limit and
compute the effective potential via the standard Cornwall--Jackiw--Tomboulis
(CJT) formalism \cite{CJT} which has the one-particle irreducible diagrams
included self-consistently. The detailed derivation of the finite $N$ case
is given in the Appendix A. We will only summarize the large $N$ result here.

In the symmetry breaking cases, we can expand shift the field $\vec{\phi}=(%
\vec{\pi},\sqrt{N}\overline{v}+\sigma )$ and expand the Lagrangian as%
\begin{eqnarray}
\mathcal{L} &=&\frac{1}{2}\left[ \left( \partial _{\mu }\vec{\pi}\right)
^{2}-m_{\pi ,0}^{2}\vec{\pi}^{2}+\left( \partial _{\mu }\sigma \right)
^{2}-m_{\sigma ,0}^{2}\sigma ^{2}\right] -NU(v)  \notag \\
&&-g_{1}\sigma ^{3}-g_{2}\sigma \vec{\pi}^{2}-\lambda _{1}\sigma
^{4}-\lambda _{2}\sigma ^{2}\vec{\pi}^{2}-\lambda _{3}\vec{\pi}^{2}\vec{\pi}%
^{2}-\lambda _{4}\sigma \ ,
\end{eqnarray}%
where 
\begin{eqnarray}
U(\overline{v}) &=&\frac{a}{2}\overline{v}^{2}+\frac{b}{4}\overline{v}^{4}+H%
\overline{v}\ ,\   \notag \\
m_{\pi ,0}^{2} &=&a+m^{2}+b\overline{v}^{2}\ ,\ \ m_{\sigma ,0}^{2}=a+3b%
\overline{v}^{2}\ ,\ \   \notag \\
g_{1} &=&g_{2}=\frac{b\overline{v}}{\sqrt{N}}\ ,\ \ \lambda _{1}=\lambda
_{3}=\frac{b}{4N}\ ,\ \   \notag \\
\lambda _{4} &=&\frac{b}{\sqrt{N}}U^{\prime }(\overline{v})\ \ .
\end{eqnarray}%
Using the result of the Appendix A, the effective potential in the CJT
formalism in the large $N$ limit reads:%
\begin{equation}
\frac{V\left( \overline{v}\right) }{N}=\frac{a}{2}\overline{v}^{2}+\frac{b}{4%
}\overline{v}^{4}+H\overline{v}+\frac{1}{2}\int_{K}\left[ \ln
P^{-1}+P_{0}^{-1}P-1\right] +\frac{b}{4}\ L_{P}^{2}+\mathcal{O}(1/N)\ ,
\end{equation}%
where $L_{P}=\int_{K}P(K,\overline{v})$ and $P(P_{0})$\ is the
full(tree-level) propagators 
\begin{equation}
P^{-1}(K,\overline{v})=-K^{2}+m_{\pi }^{2}(\overline{v})\;,\newline
\ \ P_{0}^{-1}(K,\overline{v})=-K^{2}+m_{\pi ,0}^{2}(\overline{v})\;.
\end{equation}%
The condensate $\overline{v}=v$ is determined from minimizing the effective
potential. It satisfies 
\begin{equation}
H=v\left[ a+bv^{2}+bL_{P}\right] +\mathcal{O}(1/N)\ .  \label{v below T_{c}}
\end{equation}

In case II, $v\neq 0$ below $T_{c}$, and 
\begin{eqnarray}
m_{\pi }^{2} &=&\frac{H}{v}+m^{2}+\mathcal{O}(1/N)\ ,  \notag \\
m_{\sigma }^{2} &=&m_{\sigma ,0}^{2}(v)+bL_{P}+\mathcal{O}(1/N)\   \notag \\
&=&\frac{H}{v}+2bv^{2}+\mathcal{O}(1/N).
\end{eqnarray}%
Note that, if $H=0$ and $m=0$, the goldstone bosons (the $\pi $ fields)
remain massless below some critical temperature $T_{c}$. The condensate $v$
changes continuously to zero when $T$ approaches $T_{c}$ from below showing
that it is a second-order phase transition.

Above $T_{c}$, $v=0$ and 
\begin{eqnarray}
m_{\pi }^{2} &=&a+m^{2}+bL_{P}+\mathcal{O}(1/N)\ ,  \notag \\
m_{\sigma }^{2} &=&a+bL_{P}+\mathcal{O}(1/N)\ .  \label{m above T_{c}}
\end{eqnarray}

The entropy density of the system is given by the thermal dynamical relation 
$s=-\partial V(v)/\partial T$ both below and above $T_{c}$.

\section{Shear Viscosity}

\subsection{The Boltzmann Equation}

The equations needed to compute $\eta $ in the large $N$ limit using the
Boltzmann equation are derived in the Appendix B. The $\sigma $ distribution
function is subleading, so only the $\pi $ distribution functions are
needed. All the $\pi _{i}$ components are described by the same distribution 
$f^{\pi }(\mathbf{x},\mathbf{p},t)\equiv f_{p}^{\pi }(x)$ (a function of
space, time and momentum), whose evolution is described by the Boltzmann
equation \ \ 
\begin{equation}
\frac{p^{\mu }}{E_{p}}\partial _{\mu }f_{p}^{\pi }(x)=\frac{1}{2N}\int_{123}d%
\overline{{\Gamma }}_{12;3p}^{\pi \pi \rightarrow \pi \pi }\left\{
f_{1}^{\pi }f_{2}^{\pi }F_{3}^{\pi }F_{p}^{\pi }-F_{1}^{\pi }F_{2}^{\pi
}f_{3}^{\pi }f_{p}^{\pi }\right\} ,
\end{equation}%
where $F_{i}^{\pi }\equiv 1+f_{i}^{\pi }$, $E_{p}=\sqrt{\mathbf{p}%
^{2}+m_{\pi }^{2}}$. The weighted measure is 
\begin{equation}
d\overline{\Gamma }_{12;3p}\equiv |\mathcal{T}_{12;3p}|^{2}\frac{(2\pi
)^{4}\delta ^{4}(k_{1}+k_{2}-k_{3}-p)}{2^{4}E_{1}E_{2}E_{3}E_{p}}%
\prod_{i=1}^{3}\frac{d^{3}\mathbf{k}_{i}}{(2\pi )^{3}}\ ,  \label{dGamma}
\end{equation}%
where $\mathcal{T}$ is the scattering amplitude for particles with momenta $%
1,2\rightarrow 3,p$. 
\begin{equation}
|\mathcal{T}_{12;3p}|^{2}=\left\vert 2b+\frac{4b^{2}v^{2}}{s-m_{\sigma }^{2}}%
\right\vert ^{2}+\left\vert 2b+\frac{4b^{2}v^{2}}{t-m_{\sigma }^{2}}%
\right\vert ^{2}+\left\vert 2b+\frac{4b^{2}v^{2}}{u-m_{\sigma }^{2}}%
\right\vert ^{2}.  \label{F1}
\end{equation}%
The first term is corresponding to the $s$-channel $\pi _{i}\pi
_{i}\rightarrow \pi _{j}\pi _{j}$ scattering while the second and third
terms corresponding to the $t$- and $u$-channel $\pi _{i}\pi _{j}\rightarrow
\pi _{i}\pi _{j}$ scattering. As mentioned in the introduction, the
potentially divergent $s$-channel contribution is avoided by adding the $\pi 
$ mass term such that the $\sigma $ would never be on-shell and the system
remains a perturbative one.

It is now straightforward to compute $\eta $ using the Boltzmann equation.
We follow the same procedure as in \cite{Dobado:2001jf}. It is known that
computing $\eta $ in this approach is essentially a variational problem \cite%
{Resibois}. We will go one step further to show that the procedure we take
can systematically approach the correct answer. Hence it does not rely on
the Ansatz one takes in the computation. This procedure can be used in the
computation of bulk viscosity as well.

In local thermal equilibrium, the distribution function $\overline{f}%
_{p}^{\pi }(x)=\left( e^{\beta (x)V_{\mu }(x)p^{\mu }}-1\right) ^{-1}$ with $%
\beta (x)$ the inverse temperature and $V^{\mu }(x)$ the four velocity at
the space-time point $x$. A small deviation of $f_{p}$ from local
equilibrium is parametrized as 
\begin{equation}
f_{p}^{\pi }(x)=\overline{f}_{p}^{\pi }(x)\left[ 1-\overline{F}_{p}^{\pi
}(x)\chi _{p}(x)\right] ,
\end{equation}%
where $\overline{F}_{p}^{\pi }\equiv 1+\overline{f}_{p}^{\pi }$. The energy
momentum tensor is%
\begin{equation}
T_{\mu \nu }(x)=N\int \frac{\mathrm{d}^{3}\mathbf{p}}{(2\pi )^{3}}\frac{%
p_{\mu }p_{\nu }}{E_{p}}f_{p}^{\pi }(x)\ .  \label{dT}
\end{equation}%
We will choose the $\mathbf{V}(x)=0$ frame for the point $x$. This implies $%
\partial _{\nu }V^{0}=0$ after taking a derivative on $V_{\mu }(x)V^{\mu
}(x)=1$. Furthermore, the conservation law at equilibrium $\partial _{\mu
}T^{\mu \nu }|_{\chi _{p}=0}=0$ allows us to replace $\partial _{t}\beta (x)$
and $\partial _{t}\mathbf{V}(x)$ by terms proportional to $\nabla \cdot 
\mathbf{V}(x)$ and $\mathbf{\nabla }\beta (x)$. Thus, to the first order in
a derivative expansion, $\chi _{p}(x)$ can be parametrized as 
\begin{equation}
\chi _{p}(x)=\beta (x)A(p)\nabla \cdot \mathbf{V}(x)+\beta
(x)B_{ij}(p)\nabla _{\lbrack i}V_{j]}(x)\ ,  \label{df1}
\end{equation}%
where $i$ and $j$ are spacial indexes, $B_{ij}(p)\equiv B(p)\left( \hat{p}%
_{i}\hat{p}_{j}-\frac{1}{3}\delta _{ij}\right) $ and $\nabla _{\lbrack
i}V_{j]}\equiv \left( \nabla _{i}V_{j}+\nabla _{j}V_{i}-\frac{1}{3}\delta
_{ij}\nabla \cdot \mathbf{V}(x)\right) /2$. $A$ and $B$ are functions of $x$
and $p$, but we have suppressed the $x$ dependence.

Substituting (\ref{df1}) into the Boltzmann equation, one obtains a
linearized equation for $B$%
\begin{eqnarray}
&&\left( p_{i}p_{j}-\frac{1}{3}\delta _{ij}\mathbf{p}^{2}\right)  \notag \\
&=&\frac{E_{p}}{2N}\int_{123}d\overline{\Gamma }_{12;3p}^{\pi \pi
\rightarrow \pi \pi }\overline{F}_{1}^{\pi }\overline{F}_{2}^{\pi }\overline{%
f}_{3}^{\pi }(\overline{F}_{p}^{\pi })^{-1}\left[
B_{ij}(p)+B_{ij}(k_{3})-B_{ij}(k_{2})-B_{ij}(k_{1})\right] ,  \label{Dx}
\end{eqnarray}%
where we have dropped the factor $\nabla _{\lbrack i}V_{j]}$ contracting
both sides of the equation. There is another integral equation associated
with $\nabla \cdot \mathbf{V}(x)$ which is related to the bulk viscosity $%
\zeta $ that will not be discussed in this paper. The $\mathbf{\nabla }\cdot
\beta $ and $\partial _{t}\mathbf{V}$ terms in $p^{\mu }\partial _{\mu }%
\overline{f}_{p}^{\pi }$ will cancel each other by the energy momentum
conservation in equilibrium mentioned above.

In equilibrium the energy momentum tensor depends on pressure ${\mathcal{P}}%
(x)$ and energy density $\epsilon (x)$ as $T_{\mu \nu }^{(0)}(x)=\left\{ {%
\mathcal{P}}(x)+\epsilon (x)\right\} V_{\mu }(x)V_{\nu }(x)-{\mathcal{P}}%
(x)\delta _{\mu \nu }$. A small deviation away from equilibrium gives
additional contribution to $T_{\mu \nu }$ whose spacial components define
the shear and bulk viscosity 
\begin{equation}
\delta T_{ij}=-2\eta \nabla _{\lbrack i}V_{j]}(x)+\zeta \delta _{ij}\nabla
\cdot \mathbf{V}(x)\ .
\end{equation}%
$\delta T_{ij}$ can be computed using Eq.(\ref{dT}), 
\begin{equation}
\delta T_{ij}=-N\int \frac{\mathrm{d}^{3}\mathbf{p}}{(2\pi )^{3}E_{p}}%
p^{i}p^{j}\overline{f}_{p}^{\pi }(1+\overline{f}_{p}^{\pi })\chi _{p}(x).
\end{equation}%
The above two equations imply 
\begin{eqnarray}
\eta &=&\frac{N}{10T}\int \frac{\mathrm{d}^{3}\mathbf{p}}{(2\pi )^{3}E_{p}}%
\overline{f}_{p}^{\pi }\overline{F}_{p}^{\pi }\left( p_{i}p_{j}-\frac{1}{3}%
\delta _{ij}\mathbf{p}^{2}\right) B_{ij}(p)  \notag \\
&\equiv &\left\langle S|B\right\rangle .  \label{D3x}
\end{eqnarray}%
Substituting Eq.(\ref{Dx}) into Eq.(\ref{D3x}) yields 
\begin{eqnarray}
\eta &=&\frac{1}{20T}\int_{123p}d\overline{{\Gamma }}_{12;3p}^{\pi \pi
\rightarrow \pi \pi }\overline{F}_{1}^{\pi }\overline{F}_{2}^{\pi }\overline{%
f}_{3}^{\pi }\overline{f}_{p}^{\pi }B_{ij}(p)\left[
B_{ij}(p)+B_{ij}(k_{3})-B_{ij}(k_{2})-B_{ij}(k_{1})\right]  \notag \\
&=&\frac{1}{80T}\int_{123p}d\overline{{\Gamma }}_{12;3p}^{\pi \pi
\rightarrow \pi \pi }\overline{F}_{1}^{\pi }\overline{F}_{2}^{\pi }\overline{%
f}_{3}^{\pi }\overline{f}_{p}^{\pi }\left[
B_{ij}(p)+B_{ij}(k_{3})-B_{ij}(k_{2})-B_{ij}(k_{1})\right] ^{2}  \notag \\
&\equiv &\left\langle B\left\vert C\right\vert B\right\rangle ,  \label{D5}
\end{eqnarray}%
where we have used that $\eta $ is invariant under $1\leftrightarrow 2$, $%
3\leftrightarrow p$, and $12\leftrightarrow 3p$ in the second line. It is
easy to see from the more symmetric form in the second line that $\eta $ is
non-negative so the matrix $C$ in the third line is positive definite.

\subsection{A Variational Approach}

Now we review the arguments that the computation of $\eta $ can be
formulated as a variational problem \cite{Resibois,Arnold:2003zc}. Let us
rewrite Eq.(\ref{Dx}) as%
\begin{equation}
\left\vert S\right\rangle =C\left\vert B\right\rangle ,  \label{A}
\end{equation}%
whose projection onto $\left\vert B\right\rangle $ is just 
\begin{equation}
\left\langle S|B\right\rangle =\left\langle B\left\vert C\right\vert
B\right\rangle  \label{B}
\end{equation}%
of Eqs. (\ref{D3x}) and (\ref{D5}). Technically, solving the projected
equation (\ref{B}) is easier than solving the integral equation (\ref{A}).
But this will give a wrong $\eta $. However, 
\begin{eqnarray}
\eta &=&-\left\langle B\left\vert C\right\vert B\right\rangle +2\left\langle
S|B\right\rangle  \notag \\
&=&-\left\langle B-C^{-1}S\left\vert C\right\vert B-C^{-1}S\right\rangle
+\left\langle S\left\vert C^{-1}\right\vert S\right\rangle ,
\end{eqnarray}%
where $\left\vert B-C^{-1}S\right\rangle \equiv \left\vert B\right\rangle
-C^{-1}\left\vert S\right\rangle $. Thus, if (\ref{B}) is satisfied but not (%
\ref{A}), then $\eta \leq \left\langle S\left\vert C^{-1}\right\vert
S\right\rangle $ because $C$ is positive definite. This implies that a
variational calculation of $\eta $ is possible. One just demands (\ref{B})
and try to find the Ansatz that gives the maximum $\eta $.

\subsection{Beyond Variation---Finding the Solution Systematically}

In a variational calculation, one starts with an Ansatz of $B(p)$. Assume
that $B(p)$ is a\ smooth function, one can expand it using a specific set of
orthogonal polynomials: 
\begin{equation}
B(p)=|\mathbf{p}|^{y}\sum_{r=0}^{r_{\max }}b_{r}B^{(r)}(z(p))\ .
\end{equation}%
where $z(p)=\beta p$, and $B^{(r)}(z)$ is a polynomial up to $z^{r}$ and $%
b_{r}$ is its coefficient. The overall factor $|\mathbf{p}|^{r}$ will be
chosen by trial and error to get the fastest convergence. The orthogonality
condition 
\begin{equation}
\frac{1}{15T}\int \frac{\mathrm{d}^{3}\mathbf{p}}{(2\pi )^{3}}\frac{%
\left\vert \mathbf{p}\right\vert ^{2+y}}{E_{p}}\overline{f}_{p}^{\pi }%
\overline{F}_{p}^{\pi }B^{(r)}(z)B^{(s)}(z)=\widetilde{S}^{(r)}\delta
_{r,s}\ 
\end{equation}%
can be used to construct the $B^{(r)}(z)$ polynomials up to normalization
constants. For simplicity, we will choose 
\begin{equation}
B^{(0)}(z)=1\ .
\end{equation}%
Then, Eq.(\ref{D5}) can be rewritten as 
\begin{equation}
\eta _{trial}=\left\langle b\left\vert \widetilde{C}_{r_{\max }}\right\vert
b\right\rangle \ ,  \label{H1}
\end{equation}%
where $\left\vert b\right\rangle =(b_{0},b_{1}\ldots ,b_{r_{\max }})^{T}$
and $\widetilde{C}_{r_{\max }}$ is positive definite, while Eq.(\ref{D3x})
can be rewritten as 
\begin{equation}
\eta _{trial}=\left\langle \widetilde{S}|b\right\rangle .  \label{H2}
\end{equation}%
Solving Eq.(\ref{H1}) and Eq.(\ref{H2}), we have $\left\vert b\right\rangle =%
\widetilde{C}_{r_{\max }}^{-1}\left\vert \widetilde{S}\right\rangle $, and 
\begin{equation}
\eta _{trial}=\left\langle \widetilde{S}\left\vert \widetilde{C}_{r_{\max
}}^{-1}\right\vert \widetilde{S}\right\rangle \ .
\end{equation}%
Now, according to the orthogonality condition, $\eta =\left\langle 
\widetilde{S}|b\right\rangle =Nb_{0}\widetilde{S}^{(0)}$. The other
components of $\left\vert \widetilde{S}\right\rangle $ are zero, and 
\begin{equation}
\eta _{trial}=\left( N\widetilde{S}^{(0)}\right) ^{2}\left( \widetilde{C}%
_{r_{\max }}^{-1}\right) _{00}\ .  \label{eta_trial}
\end{equation}

It can be shown that $\eta _{trial}$ increases with $r_{\max }$
monotonically. Thus, one can approach the true $\eta $ value systematically
by increasing $r_{\max }$. The proof is as follows. $\widetilde{C}_{n+1}$
denotes a $\left( n+1\right) \times \left( n+1\right) $ matrix with elements 
$\widetilde{C}_{ij}\left( i,j=0-n\right) $. Then the following identity
holds:%
\begin{equation}
\left( \widetilde{C}_{n+1}^{-1}\right) _{00}^{-1}-\left( \widetilde{C}%
_{n+2}^{-1}\right) _{00}^{-1}=\frac{\det \left( X\right) ^{2}}{\det \left(
Y\right) \det \left( Z\right) },
\end{equation}%
where $X$ is a $\left( n+1\right) \times \left( n+1\right) $ matrix with
elements $\widetilde{C}_{ij}\left( i=1-\left( n+1\right) ,j=2-(n+2)\right) $%
, $Y$ is a $n\times n$ matrix with elements $\widetilde{C}_{ij}\left(
i,j=1-n\right) $, while $Z$ is a $\left( n+1\right) \times \left( n+1\right) 
$ matrix with elements $\widetilde{C}_{ij}\left( i,j=1-\left( n+1\right)
\right) $. Both $X$ and $Y$ are positive definite, so $\det \left( Y\right)
\geq 0$ and $\det \left( Z\right) \geq 0$. Furthermore, $X$ is a real
matrix, so $\det \left( X\right) ^{2}\geq 0$. One then concludes that $%
\left( \widetilde{C}_{n+1}^{-1}\right) _{00}\leq \left( \widetilde{C}%
_{n+2}^{-1}\right) _{00}$ which implies that $\eta _{trial}$ increases with $%
r_{\max }$ monotonically.

Numerically, this algorithm converges very fast. For the case II(a) shown in
Fig. 1, using $y=1.89$, $\eta _{trial}$ increases by $\sim 0.4\%$, $\sim
0.03\%$, and $\sim 0.003\%$, when $r_{\max }$ increases from 0 to 1, 1 to 2,
and 2 to 3, respectively, for $T=30-300$.

\subsection{Numerical Results}

In the large $N$ limit, $B_{ij}=\mathcal{O}(Nb^{-2})$ and $\eta =\mathcal{O}%
(N^{2}b^{-2})$ by Eqs.(\ref{Dx}), (\ref{D5}), and $d\overline{\Gamma }=%
\mathcal{O}(b^{2})$. Combined with $s=\mathcal{O}(N)$, we have $\eta /s=%
\mathcal{O}(Nb^{-2})$. $\eta /\left( sN\right) $ is shown in Fig. 1 for
cases I-III. We see that the $\eta /s$ behavior could be different from that
of the one component scalar model \cite{Chen:2007jq}. In case I, the system
is always in the symmetric phase, and $\eta /s$ is monotonically decreasing
in $T$. In case II, the system has a second order phase transition. $\eta /s$
decreases monotonically below $T_{c}$, develops a cusp at $T_{c}$. Then,
depends on the parameters used, $\eta /s$ could be decreasing (II(a)) or
increasing (II(b)) in $T$. In II(b), $\eta /s$ does not reach a local
minimum at $T_{c}$. We will discuss this case in more details later. Case
III is similar to case II except that the cusp is smoothed out.

\begin{figure}[tbp]
\scalebox{0.5}{\includegraphics{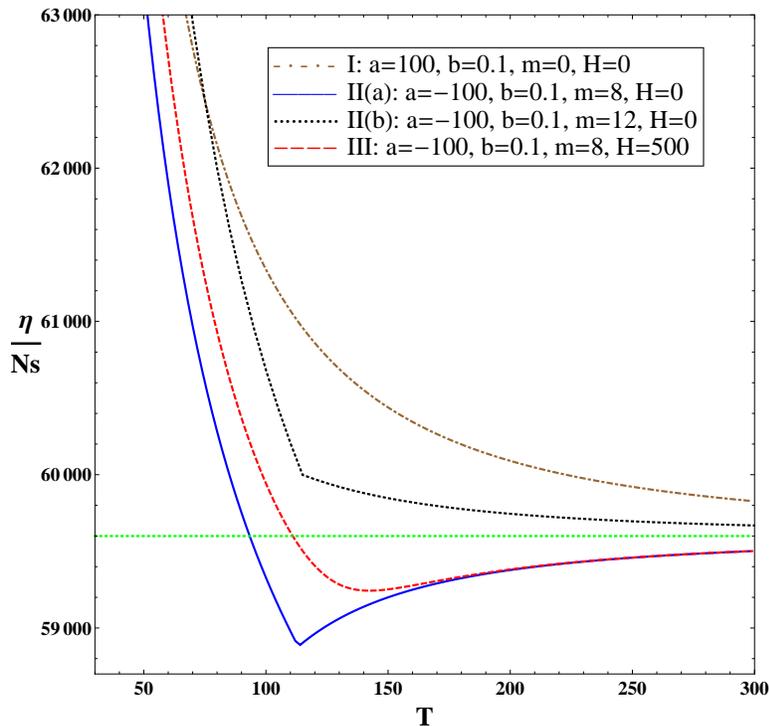}} 
\caption{$\protect\eta /(sN)$ vs. $T$ for cases with a second-order phase
transition (solid and dotted curves), a crossover (dashed curve), and with
no phase transition (dash-dotted curve). The horizontal dotted line is the
asymptotic line for $\protect\eta /(sN)$ at high $T$, which is calculated
from the massless case. Parameter can be in arbitrary units.}
\end{figure}

In cases I-III, $\eta /s$ is monotonically decreasing near $T=0$. This is
because $s$ approaches zero exponentially ($\pi $'s are massive) while $\eta 
$ approaches zero via power laws. This behavior persists even when $\pi $'s
are massless, but by a different reason. In this case, below $T_{c}$,$\
\sigma $ can be integrated out. The resulting theory is a non-linear $\sigma 
$ model with massless\ goldstone bosons (the $\pi $'s) that couple
derivatively to each other. So $\pi $'s become free particle at $T=0$, and
the interaction becomes stronger at higher $T$. Tow show this more
explicitly, when $m=H=0$, Eq.(\ref{F1}) can be recast as%
\begin{equation}
\frac{|\mathcal{T}|^{2}}{4b^{2}}=\left\vert \frac{1}{1-m_{\sigma }^{2}/s}%
\right\vert ^{2}+\left\vert \frac{1}{1-m_{\sigma }^{2}/t}\right\vert
^{2}+\left\vert \frac{1}{1-m_{\sigma }^{2}/u}\right\vert ^{2}.  \label{T}
\end{equation}%
Thus, 
\begin{equation}
|\mathcal{T}|^{2}\propto \frac{T^{2}}{m_{\sigma }^{2}},
\end{equation}%
with $m_{\sigma }$ approaching a constant at small $T$, the couplings
between pions are weaker at lower $T$ as mentioned above. And because
smaller coupling implies larger $\eta /s$, $\eta /s$ decreases monotonically
near $T=0$ for massless $\pi $'s.

As $T\rightarrow \infty $, all the systems are in the symmetric phase. The
only scales in the problem are $m_{\pi }$ and $T$. We find that $\eta /s$
has the $1/T$ expansion 
\begin{equation}
\frac{\eta }{s}=\frac{k_{0}N}{b^{2}}\left( 1+\frac{k_{1}}{\sqrt{b}}\frac{%
a+m^{2}}{T^{2}}+\cdots \right) \ ,  \label{high T}
\end{equation}%
with $k_{0}\simeq 571+80\sqrt{b}+\mathcal{O}(b)$ and $k_{1}\simeq 0.84-0.12%
\sqrt{b}+\mathcal{O}(b)$. The leading term in the expansion is the straight
line in Fig. 1 which corresponds to the $\eta /s$ for a theory with $m=a=H=0$%
. (The slow running of the coupling $b$ has been neglected. $T$ is the only
scale in the problem, so the dimensionless $\eta /s$ can only depend on the
dimensionless coupling $b$ but not $T.$) The leading $T$ dependence comes
from the $a+m^{2}$ term which has a positive sign for the symmetric phase ($%
a>0$). For the symmetric breaking phase ($a<0$), however, $a+m^{2}$ could
still be positive. Numerically, this give a $\eta /s$ which does not reach a
local minimum at $T_{c}$ for a second order phase transition as shown in
case II(b) of Fig. 1.

\begin{figure}[tbp]
\scalebox{0.7}{\includegraphics{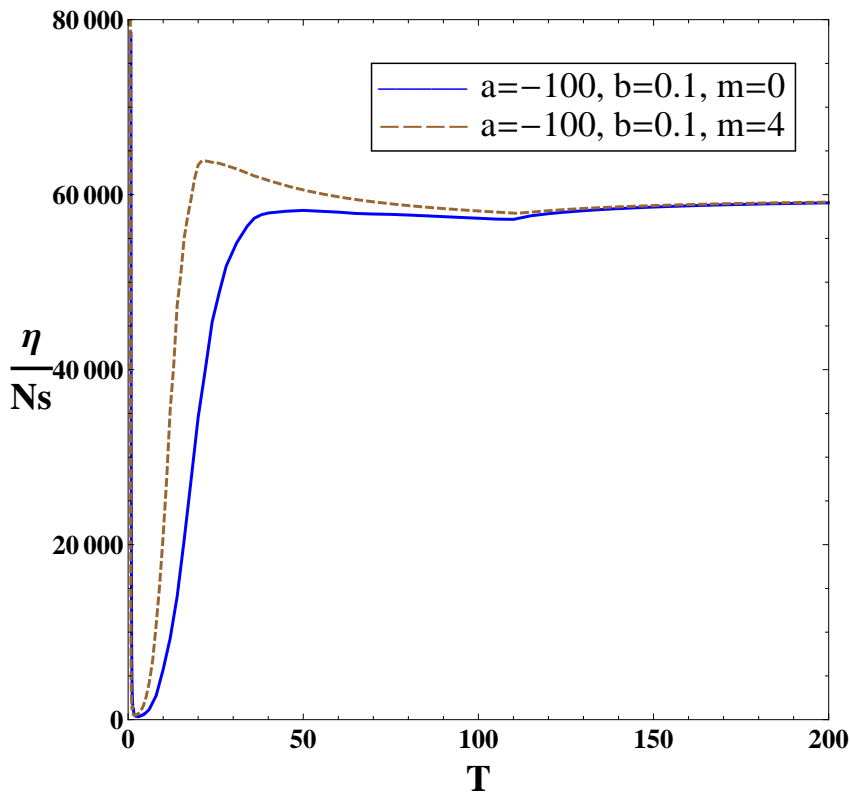}} 
\caption{$\protect\eta /(sN)$ vs. $T$ for cases where the resonance effects
are important (the thermal width of $\protect\sigma $ is included). A
\textquotedblleft double dip\textquotedblright\ behavior is seen with the $%
m=0$ case. Parameter can be in arbitrary units.}
\end{figure}

One might worry that whether case II(b) is qualified as a second order phase
transition. After all, the order parameter is defined on one component $\phi
_{N}$ whose mass is different from all the other $N-1$ components. If we
remove the $\phi _{N}$, then the system does not have a phase transition in
the first place. To answer this question, we study a similar model with just
two real scalar fields \cite{new}. One of the fields condenses below $T_{c}$
and the other stays in the symmetric phase. For simplicity, the interaction
between the two fields is turned off. It is found that the $\eta /s$
behavior in this model is similar to that of II(b).

Our results of case II(a) and II(b) differ from that of Ref. \cite%
{Dobado:2009ek} which has the minimum of $\eta /s$ below $T_{c}$. In \cite%
{Dobado:2009ek}, the parameters of the Lagrangian are tuned to mimic the $%
\pi \pi $ scattering in the real world. Thus, the S-channel $\pi \pi $
scattering diagrams are resumed to reflect the strong $\pi \pi $ scattering
in intermediate energies. In our case, we keep the coupling $b$ small and $%
2m_{\pi }>m_{\sigma }$ to make sure $\pi \pi $ scattering stays above the $%
\sigma $ resonance, such that we can apply the Boltzmann equation to compute 
$\eta $ reliably. If we set $m=0$, such that $m_{\pi }=0$ below $T_{c}$,
then the $\sigma $ can become on-shell in $\pi \pi $ scattering when $%
T\simeq \mathcal{O}(m_{\sigma }|_{T=0})\simeq b^{1/2}T_{c}<T_{c}$. Then,
indeed, a second local minimum (and sometimes also an absolute minimum,
depending on the parameters) of $\eta /s$ below $T_{c}$ can be formed to
have the \textquotedblleft double dip\textquotedblright\ structure.
Furthermore, it is conceivable that by tuning parameters, one can make the
two local minimums to be close to each other and with the cusp at $T_{c}$\
smoothed out such that one just sees the single minimum below $T_{c}$ as is
shown in \cite{Dobado:2009ek}. An $\eta /s$ plot with this feature is shown
in Fig. 2 where the thermal width of $\sigma $ \cite{Nishikawa:2003}\ is
included in the computation of $\eta $. However, the reader should be warned
that while it might a generic feature to have a dip in $\eta /s$ below $%
T_{c} $\ by a strong resonance, the $\eta $ computed with Boltzmann equation
in this case might not be reliable as discussed in Section 2.

\section{Conclusion}

We have discussed in details the computation procedure and the rich
phenomena of the $\eta /s$ behavior in weakly coupled $N$-component real
scalar field theories. We have found that $\eta /s$ can have a
\textquotedblleft double dip\textquotedblright\ behavior due to resonances
and the phase transition. It is conceivable that by tuning parameters, one
can make the two local minimums to be close to each other and with the cusp
at $T_{c}$\ smoothed out such that one just sees the single minimum below $%
T_{c}$ as is shown in \cite{Dobado:2009ek}. If an explicit goldstone mass
term is added, then $\eta /s$ can either decrease monotonically in
temperature or, as seen in many other systems, reach a minimum at the phase
transition. We have also shown how to go beyond the original variational
approach to make the Boltzmann equation computation of $\eta $ systematic.

\section{ACKNOWLEDGEMENTS}

We thank Eiji Nakano and Di-Lun Yang for involvement in the early stage of
this paper. We also thank Brian Smigielski for careful reading of the
manuscript. JWC, CTH and HHL are supported by the NSC and NCTS of Taiwan.
The work of M.H. is supported by NSFC10735040, NSFC10875134, and K.C.Wong
Education Foundation, Hong Kong.

\section{Appendix A: CJT Formalism}

In this appendix we derive the effective potential of the Lagrangian

\begin{equation}
\mathcal{L}=\frac{1}{2}(\partial _{\mu }\vec{\phi})^{2}-\overrightarrow{H}%
\cdot \vec{\phi}-\frac{1}{2}a\vec{\phi}^{2}-\frac{1}{4}b\left( \vec{\phi}%
^{2}\right) ^{2}-\frac{1}{6}c\left( \vec{\phi}^{2}\right) ^{3}\ .
\end{equation}%
The $\overrightarrow{H}\cdot \vec{\phi}$ term is included so one can mimic a
cross-over with a non-zero $\overrightarrow{H}$. The last term is a
dimension six operator whose effect to $\eta /s$ is not studied in the main
text but is included here for completeness. There could be two additional
terms with dimension six: $\vec{\phi}^{2}\left( \vec{\phi}\cdot \partial ^{2}%
\vec{\phi}\right) $, $\vec{\phi}\cdot \partial ^{2}\partial ^{2}\vec{\phi}$
(the other terms are related to these ones via integration by parts). These
terms can be removed by field redefinition or, equivalently, by applying the
equation of motion. The inclusion of the dimension six terms shows that this
is an effective field theory, which is valid under the cut-off scale $%
1/\left( N\sqrt{c}\right) $ and is renormalized order by order in the
momentum expansion $Np\sqrt{c}$, $p$ being a typical momentum scale in the
problem. $a$, $b$, and $c$ are renormalized quantities and the counterterm
Lagrangian is not shown. The renormalization condition is that the
counterterms do not change the particle mass and the four- and six-point
couplings at threshold.

We will use the standard Cornwall--Jackiw--Tomboulis (CJT) formalism \cite%
{CJT} which has the one-particle irreducible diagrams included
self-consistently. The effective potential in the CJT formalism reads:

\begin{eqnarray}
V &=&\frac{a}{2}\overline{v}^{2}+\frac{b}{4}\overline{v}^{2}+\frac{c}{6}%
\overline{v}^{6}-H\overline{v}  \notag \\
&&+\frac{1}{2}\int_{K}\left[ \ln S^{-1}+S_{0}^{-1}S-1\right] +\frac{N-1}{2}%
\int_{K}\left[ \ln P^{-1}+P_{0}^{-1}P-1\right]  \notag \\
&&+\left( N+1\right) \left( N-1\right) \left( \frac{b}{4}+\frac{c}{2}%
\overline{v}^{4}\right) L_{P}^{2}  \notag \\
&&+3\left( \frac{b}{4}+\frac{5}{2}c\overline{v}^{2}\right) L_{S}^{2}+\left(
N-1\right) \left( \frac{b}{2}+3c\overline{v}^{2}\right) L_{S}L_{P}  \notag \\
&&+\frac{c}{6}\left( N^{2}-1\right) \left( N+3\right) L_{P}^{3}+\frac{c}{2}%
\left( N^{2}-1\right) L_{P}^{2}L_{S}  \notag \\
&&+\frac{3c}{2}\left( N-1\right) L_{P}L_{S}^{2}+\frac{5c}{2}L_{S}^{3}\ ,
\end{eqnarray}%
where $L_{S}=\int_{K}\,S(K,\overline{v})$ and $L_{P}=\int_{K}\,P(K,\overline{%
v})$, and where $S(S_{0})$ and $P(P_{0})$\ are the full(tree-level)
propagators: 
\begin{eqnarray}
S^{-1}(K,\overline{v}) &=&-K^{2}+m_{\sigma }^{2}(\overline{v})\;,\newline
S_{0}^{-1}(K,\overline{v})=-K^{2}+m_{\sigma ,0}^{2}(\overline{v})\;,  \notag
\\
P^{-1}(K,\overline{v}) &=&-K^{2}+m_{\pi }^{2}(\overline{v})\;,\newline
P_{0}^{-1}(K,\overline{v})=-K^{2}+m_{\pi ,0}^{2}(\overline{v})\;,
\end{eqnarray}%
with the tree-level masses $m_{\sigma ,0}^{2}=a+3b\overline{v}^{2}+5c%
\overline{v}^{4}$ and $m_{\pi ,0}^{2}=a+b\overline{v}^{2}+c\overline{v}^{4}$%
. The expression for $V$ is consistent with that of \cite{Lenaghan:1999si}
in the $c=0$ limit.

The self-consistent one- and two-point Green's functions satisfy 
\begin{eqnarray}
\left. \frac{\delta V}{\delta \overline{v}}\right\vert _{\overline{v}%
=v,S=S(v),P=P(v)} &\equiv &0\;,\;\;\;\left. \frac{\delta V}{\delta S}%
\right\vert _{\overline{v}=v,S=S(v),P=P(v)}\equiv 0\ ,  \notag \\
\left. \frac{\delta V}{\delta P}\right\vert _{\overline{v}=v,S=S(v),P=P(v)}
&\equiv &0\ .
\end{eqnarray}%
These yield%
\begin{eqnarray}
H &=&v\left\{ a+bv^{2}+cv^{4}+\left( 3b+10cv^{2}\right) L_{S}\right.  \notag
\\
&&+\left( N-1\right) \left( b+2cv^{2}\right) L_{P}+c\left( N^{2}-1\right)
L_{P}^{2}  \notag \\
&&\left. +6c\left( N-1\right) L_{S}L_{P}+15cL_{S}^{2}\right\} \ .
\end{eqnarray}%
\begin{eqnarray}
m_{\sigma }^{2} &=&m_{\sigma ,0}^{2}+\left( 3b+30cv^{2}\right) L_{S}  \notag
\\
&&+\left( N-1\right) \left( b+6cv^{2}\right) L_{P}+c\left( N^{2}-1\right)
L_{P}^{2}  \notag \\
&&+6c\left( N-1\right) L_{S}L_{P}+15cL_{S}^{2}\ .
\end{eqnarray}%
\begin{eqnarray}
m_{\pi }^{2} &=&m_{\pi ,0}^{2}+\left( b+6cv^{2}\right) L_{S}  \notag \\
&&+\left( N+1\right) \left( b+2cv^{2}\right) L_{P}+c\left( N-1\right) \left(
N+3\right) L_{P}^{2}  \notag \\
&&+2c\left( N+1\right) L_{S}L_{P}+3cL_{S}^{2}\ .
\end{eqnarray}%
In the large $N$, i.e. $N\rightarrow \infty $, limit, a sensible scaling is
to make $m_{\sigma (\pi )}=\mathcal{O}(N^{0})$. This implies 
\begin{eqnarray}
b &=&\mathcal{O}(N^{-1})\ ,\ \ c=\mathcal{O}(N^{-2})\ ,  \notag \\
v &=&\mathcal{O}(N^{1/2})\ ,\ \ H=\mathcal{O}(N^{1/2})\ ,
\end{eqnarray}%
and $V=\mathcal{O}(N)$. This is the scaling adopted in the main text. In
this large $N$ limit, the non-tadpole type loop diagrams are subleading in
the effective potential. Thus, the Hartree approximation, which neglects the
non-tadpole type loop diagrams, gives the correct result in the large $N$
limit. For example, if $H=0$, the goldstone bosons remain massless below $%
T_{c}$ as required.

\section{Appendix B: $1/N$ Expansion of the Coupled Boltzmann Equations}

In this appendix the leading contribution to the shear viscosity of an $%
\mathcal{O}(N)$ model in the large $N$ limit is derived. We start with the
coupled Boltzmann equations:

\begin{eqnarray}
\frac{p^{\mu }}{E_{p}^{\pi _{i}}}\partial _{\mu }f_{p}^{\pi _{i}}(x) &=&%
\frac{1}{2}\int_{123}d\Gamma _{12;3p}^{\pi _{i}\pi _{i}\rightarrow \pi
_{i}\pi _{i}}\left\{ f_{1}^{\pi _{i}}f_{2}^{\pi _{i}}F_{3}^{\pi
_{i}}F_{p}^{\pi _{i}}-F_{1}^{\pi _{i}}F_{2}^{\pi _{i}}f_{3}^{\pi
_{i}}f_{p}^{\pi _{i}}\right\}  \notag \\
&&+\sum_{j\neq i}\int_{123}d\Gamma _{12;3p}^{\pi _{i}\pi _{j}\rightarrow \pi
_{i}\pi _{j}}\left\{ f_{1}^{\pi _{j}}f_{2}^{\pi _{i}}F_{3}^{\pi
_{j}}F_{p}^{\pi _{i}}-F_{1}^{\pi _{j}}F_{2}^{\pi _{i}}f_{3}^{\pi
_{j}}f_{p}^{\pi _{i}}\right\}  \notag \\
&&+\sum_{j\neq i}\frac{1}{2}\int_{123}d\Gamma _{12;3p}^{\pi _{j}\pi
_{j}\rightarrow \pi _{i}\pi _{i}}\left\{ f_{1}^{\pi _{j}}f_{2}^{\pi
_{j}}F_{3}^{\pi _{i}}F_{p}^{\pi _{i}}-F_{1}^{\pi _{j}}F_{2}^{\pi
_{j}}f_{3}^{\pi _{i}}f_{p}^{\pi _{i}}\right\}  \notag \\
&&+\frac{1}{2}\int_{123}d\Gamma _{12;3p}^{\sigma \sigma \rightarrow \pi
_{i}\pi _{i}}\left\{ f_{1}^{\sigma }f_{2}^{\sigma }F_{3}^{\pi
_{i}}F_{p}^{\pi _{i}}-F_{1}^{\sigma }F_{2}^{\sigma }f_{3}^{\pi
_{i}}f_{p}^{\pi _{i}}\right\}  \notag \\
&&+\int_{123}d\Gamma _{12;3p}^{\pi _{i}\sigma \rightarrow \sigma \pi
_{i}}\left\{ f_{1}^{\pi _{i}}f_{2}^{\sigma }F_{3}^{\sigma }F_{p}^{\pi
_{i}}-F_{1}^{\pi _{i}}F_{2}^{\sigma }f_{3}^{\sigma }f_{p}^{\pi _{i}}\right\}
\notag \\
&&+\int_{12}d\Gamma _{1;2p}^{\sigma \rightarrow \pi _{i}\pi _{i}}\left\{
f_{1}^{\sigma }F_{2}^{\pi _{i}}F_{p}^{\pi _{i}}-F_{1}^{\sigma }F_{2}^{\sigma
}f_{p}^{\pi _{i}}\right\} \ ,  \notag \\
\frac{p^{\mu }}{E_{p}^{\sigma }}\partial _{\mu }f_{p}^{\sigma }(x) &=&\frac{1%
}{2}\int_{123}d\Gamma _{12;3p}^{\sigma \sigma \rightarrow \sigma \sigma
}\left\{ f_{1}^{\sigma }f_{2}^{\sigma }F_{3}^{\sigma }F_{p}^{\sigma
}-F_{1}^{\sigma }F_{2}^{\sigma }f_{3}^{\sigma }f_{p}^{\sigma }\right\} 
\notag \\
&&+\sum_{i}\frac{1}{2}\int_{123}d\Gamma _{12;3p}^{\pi _{i}\pi
_{i}\rightarrow \sigma \sigma }\left\{ f_{1}^{\pi _{i}}f_{2}^{\pi
_{i}}F_{3}^{\sigma }F_{p}^{\sigma }-F_{1}^{\pi _{i}}F_{2}^{\pi
_{i}}f_{3}^{\sigma }f_{p}^{\sigma }\right\}  \notag \\
&&+\sum_{i}\int_{123}d\Gamma _{12;3p}^{\sigma \pi _{i}\rightarrow \pi
_{i}\sigma }\left\{ f_{1}^{\sigma }f_{2}^{\pi _{i}}F_{3}^{\pi
_{i}}F_{p}^{\sigma }-F_{1}^{\sigma }F_{2}^{\pi _{i}}f_{3}^{\pi
_{i}}f_{p}^{\sigma }\right\}  \notag \\
&&+\sum_{i}\frac{1}{2}\int_{12}d\Gamma _{12;p}^{\pi _{i}\pi _{i}\rightarrow
\sigma }\left\{ f_{1}^{\pi _{i}}f_{2}^{\pi _{i}}F_{p}^{\sigma }-F_{1}^{\pi
_{i}}F_{2}^{\pi _{i}}f_{p}^{\sigma }\right\} \ .  \label{a1}
\end{eqnarray}

The measure 
\begin{equation}
d\Gamma _{12;3p}^{\pi _{i}\sigma \rightarrow \sigma \pi _{i}}\equiv |%
\mathcal{T}_{12;3p}^{\pi _{i}\sigma \rightarrow \sigma \pi _{i}}|^{2}\frac{%
(2\pi )^{4}\delta ^{4}(k_{1}+k_{2}-k_{3}-p)}{2^{4}E_{1}^{\pi
_{i}}E_{2}^{\sigma }E_{3}^{\sigma }E_{p}^{\pi _{i}}}\prod_{i=1}^{3}\frac{%
d^{3}\mathbf{k}_{i}}{(2\pi )^{3}}\ ,
\end{equation}%
and those for the other channels are defined analogously.

Note that the $\pi _{i}$ distribution is flavor independent. Thus, in Eq. (%
\ref{a1}), there are only two independent distributions $f^{\pi }\equiv
f^{\pi _{i}}$ and $f^{\sigma }$. And the coupled Boltzmann equations can be
written as:%
\begin{eqnarray}
\frac{p^{\mu }}{E_{p}^{\pi }}\partial _{\mu }f_{p}^{\pi }(x) &=&\frac{g_{\pi
}}{2}\int_{123}d\Gamma _{12;3p}^{\pi \pi \rightarrow \pi \pi }\left\{
f_{1}^{\pi }f_{2}^{\pi }F_{3}^{\pi }F_{p}^{\pi }-F_{1}^{\pi }F_{2}^{\pi
}f_{3}^{\pi }f_{p}^{\pi }\right\}  \notag \\
&&+\frac{g_{\sigma }}{2}\int_{123}d\Gamma _{12;3p}^{\sigma \sigma
\rightarrow \pi \pi }\left\{ f_{1}^{\sigma }f_{2}^{\sigma }F^{\pi
}F_{p}^{\pi }-F_{1}^{\sigma }F_{2}^{\sigma }f_{3}^{\pi }f_{p}^{\pi }\right\}
\notag \\
&&+g_{\sigma }\int_{123}d\Gamma _{12;3p}^{\pi \sigma \rightarrow \sigma \pi
}\left\{ f_{1}^{\pi }f_{2}^{\sigma }F_{3}^{\sigma }F_{p}^{\pi }-F_{1}^{\pi
}F_{2}^{\sigma }f_{3}^{\sigma }f_{p}^{\pi }\right\}  \notag \\
&&+g_{\sigma }\int_{12}d\Gamma _{1;2p}^{\sigma \rightarrow \pi \pi }\left\{
f_{1}^{\sigma }F_{2}^{\pi }F_{p}^{\pi }-F_{1}^{\sigma }f_{2}^{\pi
}f_{p}^{\pi }\right\} \ ,  \notag \\
\frac{p^{\mu }}{E_{p}^{\sigma }}\partial _{\mu }f_{p}^{\sigma }(x) &=&\frac{%
g_{\sigma }}{2}\int_{123}d\Gamma _{12;3p}^{\sigma \sigma \rightarrow \sigma
\sigma }\left\{ f_{1}^{\sigma }f_{2}^{\sigma }F_{3}^{\sigma }F_{p}^{\sigma
}-F_{1}^{\sigma }F_{2}^{\sigma }f_{3}^{\sigma }f_{p}^{\sigma }\right\} 
\notag \\
&&+\frac{g_{\pi }}{2}\int_{123}d\Gamma _{12;3p}^{\pi \pi \rightarrow \sigma
\sigma }\left\{ f_{1}^{\pi }f_{2}^{\pi }F_{3}^{\sigma }F_{p}^{\sigma
}-F_{1}^{\pi }F_{2}^{\pi }f_{3}^{\sigma }f_{p}^{\sigma }\right\}  \notag \\
&&+g_{\pi }\int_{123}d\Gamma _{12;3p}^{\sigma \pi \rightarrow \pi \sigma
}\left\{ f_{1}^{\sigma }f_{2}^{\pi }F_{3}^{\pi }F_{p}^{\sigma
}-F_{1}^{\sigma }F_{2}^{\pi }f_{3}^{\pi }f_{p}^{\sigma }\right\}  \notag \\
&&+\frac{g_{\pi }}{2}\int_{12}d\Gamma _{12;p}^{\pi \pi \rightarrow \sigma
}\left\{ f_{1}^{\pi }f_{2}^{\pi }F_{p}^{\sigma }-F_{1}^{\pi }F_{2}^{\pi
}f_{p}^{\sigma }\right\} \ ,  \label{C1}
\end{eqnarray}%
where $g_{\pi }=N-1$ and $g_{\sigma }=1$. The scattering amplitudes
(squared) are related to those in Eq. (\ref{a1}) as%
\begin{eqnarray}
g_{\pi }\left\vert \mathcal{T}_{12;3p}^{\pi \pi \rightarrow \pi \pi
}\right\vert ^{2} &=&\left\vert \mathcal{T}_{12;3p}^{\pi _{i}\pi
_{i}\rightarrow \pi _{i}\pi _{i}}\right\vert ^{2}+2\left( g_{\pi }-1\right)
\left\vert \mathcal{T}_{12;3p}^{\pi _{i}\pi _{j}\rightarrow \pi _{i}\pi
_{j}}\right\vert ^{2}+\left( g_{\pi }-1\right) \left\vert \mathcal{T}%
_{12;3p}^{\pi _{j}\pi _{j}\rightarrow \pi _{i}\pi _{i}}\right\vert ^{2} 
\notag \\
&=&\left\vert 24\lambda _{1}+4g_{2}^{2}\left( \frac{1}{s-m_{\sigma }^{2}}+%
\frac{1}{t-m_{\sigma }^{2}}+\frac{1}{u-m_{\sigma }^{2}}\right) \right\vert
^{2}  \notag \\
&&+2\left( g_{\pi }-1\right) \left\vert 8\lambda _{3}+\frac{4g_{2}^{2}}{%
t-m_{\sigma }^{2}}\right\vert ^{2}+\left( g_{\pi }-1\right) \left\vert
8\lambda _{3}+\frac{4g_{2}^{2}}{s-m_{\sigma }^{2}}\right\vert ^{2}\   \notag
\\
&\rightarrow &N\left( 2\left\vert 8\lambda _{3}+\frac{4g_{2}^{2}}{%
t-m_{\sigma }^{2}}\right\vert ^{2}+\left\vert 8\lambda _{3}+\frac{4g_{2}^{2}%
}{s-m_{\sigma }^{2}}\right\vert ^{2}\right) =\mathcal{O}(1/N).
\end{eqnarray}%
\begin{eqnarray}
\left\vert \mathcal{T}_{12;3p}^{\sigma \sigma \rightarrow \pi \pi
}\right\vert ^{2} &=&\left\vert \mathcal{T}_{12;3p}^{\pi \pi \rightarrow
\sigma \sigma }\right\vert ^{2}=\left\vert \mathcal{T}_{12;3p}^{\sigma
\sigma \rightarrow \pi _{i}\pi _{i}}\right\vert ^{2}\   \notag \\
&=&\left\vert 4\lambda _{2}+\frac{12g_{1}g_{2}}{s-m_{\sigma }^{2}}%
+4g_{2}^{2}\left( \frac{1}{u-m_{\pi }^{2}}+\frac{1}{t-m_{\pi }^{2}}\right)
\right\vert ^{2}  \notag \\
&=&\mathcal{O}(1/N^{2})\ .
\end{eqnarray}%
\begin{eqnarray}
\left\vert \mathcal{T}_{12;3p}^{\sigma \sigma \rightarrow \sigma \sigma
}\right\vert ^{2} &=&\left\vert 24\lambda _{1}+36g_{1}^{2}\left( \frac{1}{%
s-m_{\sigma }^{2}}+\frac{1}{t-m_{\sigma }^{2}}+\frac{1}{u-m_{\sigma }^{2}}%
\right) \right\vert ^{2}  \notag \\
&=&\mathcal{O}(1/N^{2})\ .
\end{eqnarray}%
\begin{eqnarray}
\left\vert \mathcal{T}_{12;3p}^{\pi \sigma \rightarrow \sigma \pi
}\right\vert ^{2} &=&\left\vert \mathcal{T}_{12;3p}^{\sigma \pi \rightarrow
\pi \sigma }\right\vert ^{2}=\left\vert \mathcal{T}_{12;3p}^{\pi _{i}\sigma
\rightarrow \sigma \pi _{i}}\right\vert ^{2}  \notag \\
&=&\left\vert 4\lambda _{2}+\frac{12g_{1}g_{2}}{u-m_{\sigma }^{2}}%
+4g_{2}^{2}\left( \frac{1}{s-m_{\pi }^{2}}+\frac{1}{t-m_{\pi }^{2}}\right)
\right\vert ^{2}  \notag \\
&=&\mathcal{O}(1/N^{2})\ .
\end{eqnarray}%
\begin{equation}
\left\vert \mathcal{T}_{1;2p}^{\sigma \rightarrow \pi \pi }\right\vert ^{2}\
=\left\vert 2g_{2}\right\vert ^{2}=\mathcal{O}(1/N)\ .
\end{equation}%
In the large $N$ limit, Eq.(\ref{C1}) is simplified to 
\begin{eqnarray}
\frac{p^{\mu }}{E_{p}^{\pi }}\partial _{\mu }f_{p}^{\pi }(x) &=&\frac{1}{2N}%
\int_{123}d\overline{\Gamma }_{12;3p}^{\pi \pi \rightarrow \pi \pi }\left\{
f_{1}^{\pi }f_{2}^{\pi }F_{3}^{\pi }F_{p}^{\pi }-F_{1}^{\pi }F_{2}^{\pi
}f_{3}^{\pi }f_{p}^{\pi }\right\}  \notag \\
&&+\frac{1}{N}\int_{12}d\overline{\Gamma }_{1;2p}^{\sigma \rightarrow \pi
\pi }\left\{ f_{1}^{\sigma }F_{2}^{\pi }F_{p}^{\pi }-F_{1}^{\sigma
}f_{2}^{\pi }f_{p}^{\pi }\right\} \ ,  \label{B1} \\
\frac{p^{\mu }}{E_{p}^{\sigma }}\partial _{\mu }f_{p}^{\sigma }(x) &=&+\frac{%
1}{2}\int_{12}d\overline{\Gamma }_{12;p}^{\pi \pi \rightarrow \sigma
}\left\{ f_{1}^{\pi }f_{2}^{\pi }F_{p}^{\sigma }-F_{1}^{\pi }F_{2}^{\pi
}f_{p}^{\sigma }\right\} ,  \label{B2}
\end{eqnarray}%
where the $N$ dependence in $d\Gamma $ is factored out already, so all the $%
N $ dependence is in the prefactors. The above equations imply 
\begin{eqnarray}
&&\left( p_{i}p_{j}-\frac{1}{3}\delta _{ij}\mathbf{p}^{2}\right)  \notag \\
&=&\frac{E_{p}^{\pi }}{2N}\int_{123}d\overline{\Gamma }_{12;3p}^{\pi \pi
\rightarrow \pi \pi }\overline{F}_{1}^{\pi }\overline{F}_{2}^{\pi }\overline{%
f}_{3}^{\pi }(\overline{F}_{p}^{\pi })^{-1}\left[ B_{ij}^{\pi
}(p)+B_{ij}^{\pi }(k_{3})-B_{ij}^{\pi }(k_{2})-B_{ij}^{\pi }(k_{1})\right] 
\notag \\
&&+\frac{1}{N}E_{p}^{\pi }\int_{12}d\overline{\Gamma }_{1;2p}^{\sigma
\rightarrow \pi \pi }\overline{F}_{1}^{\sigma }\overline{f}_{2}^{\pi }(%
\overline{F}_{p}^{\pi })^{-1}\left[ B_{ij}^{\pi }(p)+B_{ij}^{\pi
}(k_{2})-B_{ij}^{\sigma }(k_{1})\right] ,  \label{X1} \\
&&\left( p_{i}p_{j}-\frac{1}{3}\delta _{ij}\mathbf{p}^{2}\right)  \notag \\
&=&\frac{E_{p}^{\sigma }}{2}\int_{12}d\overline{\Gamma }_{12;p}^{\pi \pi
\rightarrow \sigma }\overline{F}_{1}^{\pi }\overline{F}_{2}^{\pi }(\overline{%
F}_{p}^{\sigma })^{-1}\left[ B_{ij}^{\sigma }(p)-B_{ij}^{\pi
}(k_{2})-B_{ij}^{\pi }(k_{1})\right] ,  \label{X2}
\end{eqnarray}%
where $B_{ij}$, defined in Eq.(\ref{df1}), describes how $\overline{f}$
changes when the velocity distribution is non-uniform. Eq.(\ref{X1}) demands 
$B_{ij}^{\pi }(p)=\mathcal{O}(N).$ Eq.(\ref{X2}) demands $B_{ij}^{\sigma
}(p)=\mathcal{O}(N)$ such that Eq.(\ref{X2}) remains $\mathcal{O}(1)$. Now,

\begin{eqnarray}
\eta &=&NL^{\pi }\left[ B^{\pi }\right] +L^{\sigma }\left[ B^{\sigma }\right]
\ ,  \notag \\
L^{l}\left[ B^{l}\right] &=&\frac{\beta }{15}\int \frac{\mathrm{d}^{3}%
\mathbf{p}\,\mathbf{p}^{2}}{(2\pi )^{3}E_{p}^{k}}\overline{f}_{p}^{l}%
\overline{F}_{p}^{l}B^{l}(p)\   \label{D4}
\end{eqnarray}%
In the large $N$ limit%
\begin{eqnarray}
\eta &=&NL^{\pi }\left[ B^{\pi }\right] \   \notag \\
&\simeq &NL^{\pi }\left[ B^{\pi }\right] +cL^{\sigma }\left[ B^{\sigma }%
\right] ,  \label{X}
\end{eqnarray}%
where we have added a subleading term with prefactor $c\sim \mathcal{O}%
(N^{0})$. The final result for $\eta $ should not depend on the choice of $c$%
.

Substituting Eqs.(\ref{X1},\ref{X2}) into Eq.(\ref{X}), one obtains

\begin{eqnarray}
\eta &=&\frac{\beta }{20}\int_{123p}d\overline{\Gamma }_{12;3p}^{\pi \pi
\rightarrow \pi \pi }\overline{F}_{1}^{\pi }\overline{F}_{2}^{\pi }\overline{%
f}_{3}^{\pi }\overline{f}_{p}^{\pi }B_{ij}^{\pi }(p)\left[ B_{ij}^{\pi
}(p)+B_{ij}^{\pi }(k_{3})-B_{ij}^{\pi }(k_{2})-B_{ij}^{\pi }(k_{1})\right] 
\notag \\
&&+\frac{\beta }{10}\int_{12p}d\overline{\Gamma }_{1;2p}^{\sigma \rightarrow
\pi \pi }\overline{F}_{1}^{\sigma }\overline{f}_{2}^{\pi }\overline{f}%
_{p}^{\pi }B_{ij}^{\pi }(p)\left[ B_{ij}^{\pi }(p)+B_{ij}^{\pi
}(k_{2})-B_{ij}^{\sigma }(k_{1})\right]  \notag \\
&&+\frac{\beta c}{20}\int_{12p}d\overline{\Gamma }_{12;p}^{\pi \pi
\rightarrow \sigma }\overline{F}_{1}^{\pi }\overline{F}_{2}^{\pi }\overline{f%
}_{p}^{\sigma }B_{ij}^{\sigma }(p)\left[ B_{ij}^{\sigma }(p)-B_{ij}^{\pi
}(k_{2})-B_{ij}^{\pi }(k_{1})\right] .
\end{eqnarray}%
Symmetries of the equations further gives 
\begin{eqnarray}
\eta &=&\frac{\beta }{80}\int_{123p}d\overline{\Gamma }_{12;3p}^{\pi \pi
\rightarrow \pi \pi }\overline{F}_{1}^{\pi }\overline{F}_{2}^{\pi }\overline{%
f}_{3}^{\pi }\overline{f}_{p}^{\pi }\left[ B_{ij}^{\pi }(p)+B_{ij}^{\pi
}(k_{3})-B_{ij}^{\pi }(k_{2})-B_{ij}^{\pi }(k_{1})\right] ^{2}  \notag \\
&&+\frac{\left( 2+c\right) \beta }{60}\int_{12p}d\overline{\Gamma }%
_{1;2p}^{\sigma \rightarrow \pi \pi }\overline{F}_{1}^{\sigma }\overline{f}%
_{2}^{\pi }\overline{f}_{p}^{\pi }\left[ B_{ij}^{\pi }(p)+B_{ij}^{\pi
}(k_{2})-B_{ij}^{\sigma }(k_{1})\right] ^{2}.
\end{eqnarray}%
By choosing $c=-2$, the subleading contribution can be subtracted. Thus, $%
\overline{f}^{\sigma }$and $B_{ij}^{\sigma }$ decouple from $\eta $\ and the 
$\sigma $ contribution only appears in the intermediate states of $\pi \pi $
scattering.

In summery, in the large $N$ limit, one can use 
\begin{eqnarray}
\eta &=&\frac{N\beta }{10}\int \frac{\mathrm{d}^{3}\mathbf{p}}{(2\pi
)^{3}E_{p}}\overline{f}_{p}^{\pi }\overline{F}_{p}^{\pi }\left( p_{i}p_{j}-%
\frac{1}{3}\delta _{ij}\mathbf{p}^{2}\right) B_{ij}(p)  \notag \\
&=&\frac{\beta }{80}\int_{123p}d\overline{\Gamma }_{12;3p}^{\pi \pi
\rightarrow \pi \pi }\overline{F}_{1}^{\pi }\overline{F}_{2}^{\pi }\overline{%
f}_{3}^{\pi }\overline{f}_{p}^{\pi }\left[ B_{ij}^{\pi }(p)+B_{ij}^{\pi
}(k_{3})-B_{ij}^{\pi }(k_{2})-B_{ij}^{\pi }(k_{1})\right] ^{2}\text{,}
\end{eqnarray}%
to solve $B_{ij}^{\pi }$ and $\eta $.


\begin{thebibliography}{99}
\bibitem{Jeon} S. Jeon, Phys. Rev. D \textbf{52}, 3591 (1995); S. Jeon and
L. Yaffe, Phys. Rev. D \textbf{53}, 5799 (1996).

\bibitem{KOVT1} P. Kovtun, D.T. Son, and A.O. Starinets, Phys. Rev. Lett. 
\textbf{94},111601 (2005).

\bibitem{Policastro:2001yc} G.~Policastro, D.T.~Son, and A.O.~Starinets, 
Phys.\ Rev.\ Lett.\ \textbf{87}, 081601 (2001). 

\bibitem{Policastro:2002se} G.~Policastro, D.~T.~Son and A.~O.~Starinets, 
JHEP \textbf{0209}, 043 (2002). 

\bibitem{Herzog:2002fn} C.P.~Herzog, 
J.\ High Energy Phys.\ \textbf{0212}, 026 (2002). 

\bibitem{Buchel:2003tz} A.~Buchel and J.T.~Liu, 
Phys.\ Rev.\ Lett.\ \textbf{93}, 090602 (2004). 

\bibitem{Son:2007vk} D.~T.~Son and A.~O.~Starinets, 
Ann.\ Rev.\ Nucl.\ Part.\ Sci.\ \textbf{57}, 95 (2007) [arXiv:0704.0240
[hep-th]]. 

\bibitem{Kapusta:2008vb} J.~I.~Kapusta, 
arXiv:0809.3746 [nucl-th]. 


\bibitem{Schafer:2009dj} T.~Schafer and D.~Teaney, 
Rept.\ Prog.\ Phys.\ \textbf{72}, 126001 (2009) [arXiv:0904.3107 [hep-ph]]. 


\bibitem{Jakovac:2009xn} A. Jakovac, Phys. Rev. D81: 045020 (2010).

\bibitem{Cohen:2007qr} T.~D.~Cohen, 
Phys.\ Rev.\ Lett.\ \textbf{99}, 021602 (2007) [arXiv:hep-th/0702136]. 


\bibitem{Cherman:2007fj} A.~Cherman, T.~D.~Cohen and P.~M.~Hohler, 
JHEP \textbf{0802}, 026 (2008) [arXiv:0708.4201 [hep-th]]. 




\bibitem{Kats:2007mq} Y.~Kats and P.~Petrov, 
JHEP \textbf{0901}, 044 (2009) [arXiv:0712.0743 [hep-th]]. 


\bibitem{Brigante:2007nu} M.~Brigante, H.~Liu, R.~C.~Myers, S.~Shenker and
S.~Yaida, 
Phys.\ Rev.\ D \textbf{77}, 126006 (2008) [arXiv:0712.0805 [hep-th]]. 


\bibitem{Brigante:2008gz} M.~Brigante, H.~Liu, R.~C.~Myers, S.~Shenker and
S.~Yaida, 
Phys.\ Rev.\ Lett.\ \textbf{100}, 191601 (2008) [arXiv:0802.3318 [hep-th]]. 


\bibitem{Buchel:2008vz} A.~Buchel, R.~C.~Myers and A.~Sinha, 
JHEP \textbf{0903}, 084 (2009) [arXiv:0812.2521 [hep-th]]. 


\bibitem{Gyulassy:2004zy} M.~Gyulassy and L.~McLerran, Nucl.\ Phys.\ A 
\textbf{750}, 30 (2005) {[}arXiv:nucl-th/0405013{]}. 

\bibitem{Shuryak:2004cy} E.~V.~Shuryak, Nucl.\ Phys.\ A \textbf{750}, 64
(2005) {[}arXiv:hep-ph/0405066{]}. 

\bibitem{Stoecker:2004qu} H.~Stoecker, Nucl.\ Phys.\ A \textbf{750}, 121
(2005) {[}arXiv:nucl-th/0406018{]}. 

\bibitem{Jacobs:2004qv} P.~Jacobs and X.~N.~Wang, Prog.\ Part.\ Nucl.\
Phys.\ \textbf{54}, 443 (2005) {[}arXiv:hep-ph/0405125{]}. 

\bibitem{RHIC} I.~Arsene \emph{et al.}, 
Nucl.\ Phys.\ A \textbf{757}, 1 (2005); 
B.~B.~Back \emph{et al.}, 
\emph{ibid.} \textbf{757}, 28 (2005); 
J.~Adams \emph{et al.}, 
\emph{ibid.} \textbf{757}, 102 (2005); K.~Adcox \emph{et al.}, 
\emph{ibid.} \textbf{757}, 184 (2005). 

\bibitem{Huovinen:2001cy} P.~Huovinen, P.~F.~Kolb, U.~W.~Heinz,
P.~V.~Ruuskanen and S.~A.~Voloshin, Phys.\ Lett.\ B \textbf{503}, 58 (2001) {%
[}arXiv:hep-ph/0101136{]}. 

\bibitem{Teaney:2000cw} D.~Teaney, J.~Lauret and E.~V.~Shuryak, Phys.\ Rev.\
Lett.\ \textbf{86}, 4783 (2001) {[}arXiv:nucl-th/0011058{]}. 

\bibitem{Muronga:2004sf} A.~Muronga and D.~H.~Rischke,
arXiv:nucl-th/0407114. 

\bibitem{Heinz:2005bw} U.~W.~Heinz, H.~Song and A.~K.~Chaudhuri, Phys.\
Rev.\ C \textbf{73}, 034904 (2006) {[}arXiv:nucl-th/0510014{]}. 

\bibitem{Romatschke:2007mq} P.~Romatschke and U.~Romatschke, Phys.\ Rev.\
Lett.\ \textbf{99}, 172301 (2007) {[}arXiv:0706.1522 {[}nucl-th{]}{]}. 

\bibitem{Hirano:2002ds} T.~Hirano and K.~Tsuda, Phys.\ Rev.\ C \textbf{66},
054905 (2002) {[}arXiv:nucl-th/0205043{]}. 

\bibitem{Molnar:2001ux} D.~Molnar and M.~Gyulassy, 
Nucl.\ Phys.\ A \textbf{697}, 495 (2002) [Erratum \emph{ibid.} \textbf{703},
893 (2002)]. 

\bibitem{Teaney:2003pb} D.~Teaney, 
Phys.\ Rev.\ C \textbf{68}, 034913 (2003). 

\bibitem{Luzum:2008cw} M.~Luzum and P.~Romatschke, 
Phys.\ Rev.\ C \textbf{78}, 034915 (2008) [arXiv:0804.4015 [nucl-th]]. 

\bibitem{Song:2008hj} H.~Song and U.~W.~Heinz, 
J.\ Phys.\ G \textbf{36}, 064033 (2009) [arXiv:0812.4274 [nucl-th]]. 

\bibitem{etas-gluon-lat} 
H.~B.~Meyer, 
Phys.\ Rev.\ D \textbf{76}, 101701 (2007), arXiv:0704.1801 [hep-lat]. 


\bibitem{Schafer} T. Schafer, Phys. Rev. A 76, 063618 (2007).

\bibitem{Turlapov} A. Turlapov, J. Kinast, B. Clancy, L. Luo, J. Joseph, and
J. E. Thomas, J. Low Temp. Phys. 150, 567 (2008).

\bibitem{Clancy} B. Clancy, L. Luo, J. E. Thomas Phys. Rev. Lett. 99 140401
(2007) [arXiv:0705.2782 [condmat.other]].

\bibitem{Thomas} J.E. Thomas, Nucl. Phys. A 830, 665c (2009).

\bibitem{Schaefer2} T. Schaefer, C. Chafin, e-Print: arXiv:0912.4236
[cond-mat.quant-gas]; T. Schaefer, e-Print: arXiv:1008.3876
[cond-mat.quant-gas].

\bibitem{Csernai:2006zz} L.~P.~Csernai, J.~I.~Kapusta and L.~D.~McLerran, 
Phys.\ Rev.\ Lett.\ \textbf{97}, 152303 (2006).

\bibitem{Chen:2006iga} J.~W.~Chen and E.~Nakano, 
Phys.\ Lett.\ B \textbf{647}, 371 (2007). 

\bibitem{Chen:2007jq} J.~W.~Chen, M.~Huang, Y.~H.~Li, E.~Nakano and
D.~L.~Yang, 
Phys.\ Lett.\ B \textbf{670}, 18 (2008) [arXiv:0709.3434 [hep-ph]]. 

\bibitem{Aarts:2004sd} G.~Aarts and J.~M.~Martinez Resco, Phys. Rev. D 68,
085009 (2003); 
JHEP \textbf{0402}, 061 (2004). 

\bibitem{Moore:2007ib} G.~D.~Moore, 
Phys. Rev. D76: 107702, 2007. 

\bibitem{Dobado:2009ek} A.~Dobado, F.~J.~Llanes-Estrada and
J.~M.~Torres-Rincon, 
Phys.\ Rev.\ D \textbf{80}, 114015 (2009) [arXiv:0907.5483 [hep-ph]]. 



\bibitem{Chakraborty:2010fr} P. Chakraborty, J.I. Kapusta, arXiv:1006.0257
[nucl-th]

\bibitem{Arnold:2003zc} P.~Arnold, G.~D.~Moore and L.~G.~Yaffe, JHEP \textbf{%
0305}, 051 (2003) {[}arXiv:hep-ph/0302165{]}; P.~Arnold, 
Int.\ J.\ Mod.\ Phys.\ E \textbf{16}, 2555 (2007) [arXiv:0708.0812
[hep-ph]]. 

\bibitem{Xu:2007ns} Z.~Xu and C.~Greiner, Phys.\ Rev.\ Lett.\ \textbf{100},
172301 (2008) {[}arXiv:0710.5719 {[}nucl-th{]}{]}. 


\bibitem{Chen:2009sm} J.~W.~Chen, H.~Dong, K.~Ohnishi and Q.~Wang, 
Phys.\ Lett.\ B \textbf{685}, 277 (2010) [arXiv:0907.2486 [nucl-th]]. 

\bibitem{Prakash:1993bt} M.~Prakash, M.~Prakash, R.~Venugopalan and
G.~Welke, 
Phys.\ Rept.\ \textbf{227}, 321 (1993). 


\bibitem{Dobado:2003wr} A.~Dobado and F.~J.~Llanes-Estrada, 
Phys.\ Rev.\ D \textbf{69} (2004) 116004 [arXiv:hep-ph/0309324].

\bibitem{Dobado:2001jf} A.~Dobado and S.~N.~Santalla, Phys.\ Rev.\ D \textbf{%
65}, 096011 (2002) {[}arXiv:hep-ph/0112299{]}.

\bibitem{Itakura:2007mx} K.~Itakura, O.~Morimatsu and H.~Otomo, 
Phys.\ Rev.\ D \textbf{77}, 014014 (2008) [arXiv:0711.1034 [hep-ph]]. 


\bibitem{Chen:2007xe} J.~W.~Chen, Y.~H.~Li, Y.~F.~Liu and E.~Nakano, Phys.
Rev. D76, 114011(2007).%

\bibitem{etas-supfluid} 
T.~Schafer, 
arXiv:cond-mat/0701251; 
G.~Rupak and T.~Schafer, 
arXiv:0707.1520 [cond-mat.other]. 

\bibitem{webbook} E.W.~Lemmon \emph{et al.}, Thermophysical Properties of
Fluid Systems, in \textit{NIST Chemistry WebBook}, NIST Standard Reference
Database Number 69, Eds. Linstrom P.G. \& Mallard, W.G., March 2003
(http://webbook.nist.gov).

\bibitem{Lacey:2006bc} R.~A.~Lacey \textit{et al.}, 
Phys.\ Rev.\ Lett.\ \textbf{98}, 092301 (2007); arXiv:0708.3512. 

\bibitem{Gagnon:2007qt} J.~S.~Gagnon and S.~Jeon, 
Phys.\ Rev.\ D \textbf{76}, 105019 (2007) [arXiv:0708.1631 [hep-ph]]. 

\bibitem{CJT} J.M.\ Cornwall, R.\ Jackiw, and E.\ Tomboulis, Phys.\ Rev.\ D 
\textbf{10}, 2428 (1974).

\bibitem{Resibois} P.~R\'{e}sibois and M.~d. Leener, \emph{Classical Kinetic
Theory of Fluids} (John Wiley $\&$ Sons, 1977).

\bibitem{new} J.W. Chen, C.T. Hsieh, H.H. Lin, \emph{in preparation}.

\bibitem{Nishikawa:2003} T.~Nishikawa, O.~Morimatsu, and Y.~Hidaka, 
Phys.\ Rev.\ D \textbf{68}, 076002 (2003) [arXiv:hep-ph/0302098].

\bibitem{Lenaghan:1999si} J.~T.~Lenaghan and D.~H.~Rischke, 
J.\ Phys.\ G \textbf{26}, 431 (2000) [arXiv:nucl-th/9901049]. 
\end{thebibliography}
\end{document}